\documentclass[apj]{emulateapj} 
\usepackage[colorlinks=true,urlcolor=blue,citecolor=blue,linkcolor=blue]{hyperref}

\def\cf{cf.}
\def\radyn{{ RADYN}}
\def\nicole{{ NICOLE}}
\newcommand{\be}{\begin{equation}}
\newcommand{\ee}{\end{equation}}
\def\kms{km\,s$^{-1}$}

\def\Halpha{\mbox{H\hspace{0.1ex}$\alpha$}}
\def\rmit#1{{\it #1}}
\def\eg{\rmit{e.g.,}}

\def\edt#1{{#1}}

\def\h2{\ensuremath{\mathrm{H}_2}}

\def\Ca{Ca}  

\def\CaII{\ion{Ca}{2}}
\def\Cair{\CaII\ 854.2 nm}
\def\RH{{RH}}
\def\kms{\mbox{{km s $^{-1}$}}}

\begin{document}

\title{The effect of isotopic splitting on the bisector and inversions of the solar Ca II 854.2 nm line}

   \author{Jorrit Leenaarts$^{1,2}$}\email{jorritl@astro.uio.no}
   \author{Jaime de la Cruz Rodr\'iguez$^{2}$}\email{jaime@astro.su.se}
    \author{Oleg Kochukhov$^{3}$}\email{oleg.kochukhov@physics.uu.se}
   \author{Mats Carlsson$^{1}$}\email{mats.carlsson@astro.uio.no}

\affil{$^1$ Institute of
  Theoretical Astrophysics, University of Oslo, P.O. Box 1029
  Blindern, N--0315 Oslo, Norway}
\affil{$^2$ Institute for Solar Physics, Department of Astronomy, Stockholm University,
AlbaNova University Centre, SE-106 91 Stockholm Sweden}
\affil{$^3$ Department of Physics and Astronomy, Uppsala University, Box 516, SE-75120 Uppsala, Sweden}

 \date{Received; accepted}

\begin{abstract}
The Ca II 854.2 nm spectral line is a common diagnostic of the solar
chromosphere. The average line profile shows an asymmetric core, and
its bisector shows a characteristic inverse-C shape. The line actually
consists of six components with slightly different wavelengths
depending on the isotope of calcium. This isotopic splitting of the
line has been taken into account in studies of non-solar stars, but
never for the Sun. We performed non-LTE radiative transfer
computations from three models of the solar atmosphere and show that
the asymmetric line-core and inverse C-shape of the bisector of the
854.2 nm line can be explained by isotopic splitting. We confirm this
finding by analysing observations and showing that the line asymmetry
is present irrespective of conditions in the solar
atmosphere. Finally, we show that inversions based on the
Ca~II~854.2~nm line should take the isotopic splitting into account,
otherwise the inferred atmospheres will contain erroneous velocity
gradients and temperatures.
 \end{abstract}
          
   \keywords{Sun: atmosphere --- Sun: chromosphere --- radiative transfer}
  
\section{Introduction}                          \label{sec:introduction}

The triplet of lines of \CaII\ at 849.8, 854.2 and 866.2~nm in the
solar spectrum are formed in the chromosphere of the Sun. They are
common diagnostics of the chromosphere.
%
%
%
%
The infrared triplet is sensitive to magnetic fields and is used
to infer properties of the chromospheric magnetic field based on
observations \edt{of the Stokes vector}
\citep[\eg][]{2000Sci...288.1396S,%
2001ApJ...552..871L,%
2006SoPh..235...55S,%
2010ApJ...710.1486J,%
2011A&A...527L...8D}.

All three lines show a marked asymmetry in their line cores in spatially and temporally averaged profiles
\citep{1984SoPh...90..205N},
and their bisector shows an inverse-C shape
\citep{2006ApJ...639..516U}.
The \Halpha\ line, which forms at similar heights in the chromosphere
does not show such asymmetry
\citep{2013SoPh..288...89C}.

The variation of the bisector of the \Cair\ line during the solar
cycle has been investigated by
\citet{2011ApJ...736..114P},
who found that the asymmetry of the line core, and thus the bisector,
change in phase with the solar magnetic cycle. In a follow-up study,
\citet{2013ApJ...764..153P}
found that the bisector shape in areas with high magnetic flux is
different from the bisector in quiet Sun, providing a natural
explanation for the asymmetry-activity correlation.

The actual physical mechanism that causes the solar line asymmetry has
so far not been identified.  Previous modeling efforts did not
reproduce the observed inverse-C shaped bisector
\citep{2006ApJ...639..516U,2009ApJ...694L.128L}.
These studies ignored the minority isotopes of calcium and instead assumed all calcium in the form of $^{40}$Ca, the most abundant isotope 
\citep[96.94\% in the standard solar system composition,][]{1989GeCoA..53..197A}

In the stellar community it has however been realised that the other
isotopes should be taken into account in order to explain observed
line shapes in chemically peculiar stars
%
\citep[\eg][]{2004A&A...421L...1C,2005A&A...432L..21C,2007MNRAS.377.1579C,2008A&A...480..811R}. 

With this letter we draw the attention of the solar community to the
importance of isotopic splitting in the modelling of the \CaII\ IR
triplet. Using appropriate chromospheric models, we show that isotopic
splitting can explain the observed inverse-C shaped bisector of the Ca
IR triplet lines. In addition we demonstrate that neglecting the
splitting in inversions of observed Stokes profiles can lead to
erroneous derivation of the velocity, temperature, and possibly,
magnetic field strength.

\section{Simulations and radiative transfer} \label{sec:simulations}

\begin{table}
\caption{Ca isotopes, their abundance and 3d $^2$D$_{5/2}$ -- 4p $^2$P$_{3/2}$ line wavelengths
\label{table:ca-table}}
\centering
\begin{tabular}{c c c}
\hline\hline 
Isotope & Abundance & $\lambda$ (nm)
\\
\hline 
$^{40}$Ca & 6.33 & 854.20857 \\
$^{42}$Ca & 4.15 & 854.21426 \\
$^{43}$Ca & 3.47 & 854.21696 \\
$^{44}$Ca & 4.66 & 854.21952 \\
$^{46}$Ca & 1.94 & 854.22433 \\
$^{48}$Ca & 3.61 & 854.22871 \\

\hline
\end{tabular} 
\end{table}

We constructed 5-level-plus-continuum model atoms for the 4 stable
isotopes and two extremely long-lived isotopes of calcium. We computed
the energy levels of the \mbox{4p $^2$P$_{3/2,5/2}$} and \mbox{3d $^2$D$_{1/2,3/2}$}  states using the experimental data from
  \citet{1992PhRvA..45.4675M}
and 
  \citet{1998EPJD....2...33N}.
There is a near-linear relation between the number of nucleons and the
wavelength of each line in the IR triplet; we used this relation to
extrapolate the energy levels for isotopes that were not measured. \edt{One isotope, $^{43}$Ca, shows hyperfine splitting
\citep{2011PhRvA..83a2503S}.
We do not take this into account. Because the spread of this splitting is only 0.88~pm, this does not significantly influence our results.}

We took the total calcium abundance from
\citet{2009ARA&A..47..481A}
and the relative abundance of each isotope from 
\citet{1989GeCoA..53..197A}.
In Table~\ref{table:ca-table} we give the isotopes, their abundance in
the standard $A_\mathrm{H}=12$ scale and the wavelength of the \Cair\ line.

We used three model atmospheres.
The first is the 1D semi-empirical FAL model C of
\citet{1993ApJ...406..319F}. 
The second is a time series of 240 snapshots taken at 10 s intervals computed with the 1D radiation-hydrodynamics code \radyn\
\citep[\eg][]{1992ApJ...397L..59C, 1997ApJ...481..500C}.
The third is a snapshot of a 3D radiation-MHD simulation performed with the Bifrost code
\citep{2011A&A...531A.154G}.
We chose the same snapshot as was used by
\citet{2012ApJ...749..136L}
to investigate \Halpha\ line formation, and refer to that paper for
details of the simulation.

We performed the non-LTE radiative transfer computations using the \RH\ code by 
\citet{2001ApJ...557..389U}.
For each atmospheric model we performed two computations: one where we
assumed all calcium to be in the form of $^{40}$\Ca, and one where we
included all isotopes given in Table~\ref{table:ca-table}. In the
multi-isotope computation we included the isotopes as separate atoms
that were treated simultaneously in non-LTE. FALC included a
microturbulence that varies with height as given in
\citet{1981ApJS...45..635V}.
In the RADYN simulation we used a constant microturbulence of 2
\kms. The 3D Bifrost snapshot did not include microturbulence. Each
column in this model was treated independently as a 1D plane-parallel
atmosphere.

\section{Results} \label{sec:results}

\subsection{Profile asymmetry and bisector}

\begin{figure}
  \includegraphics[width=\columnwidth]{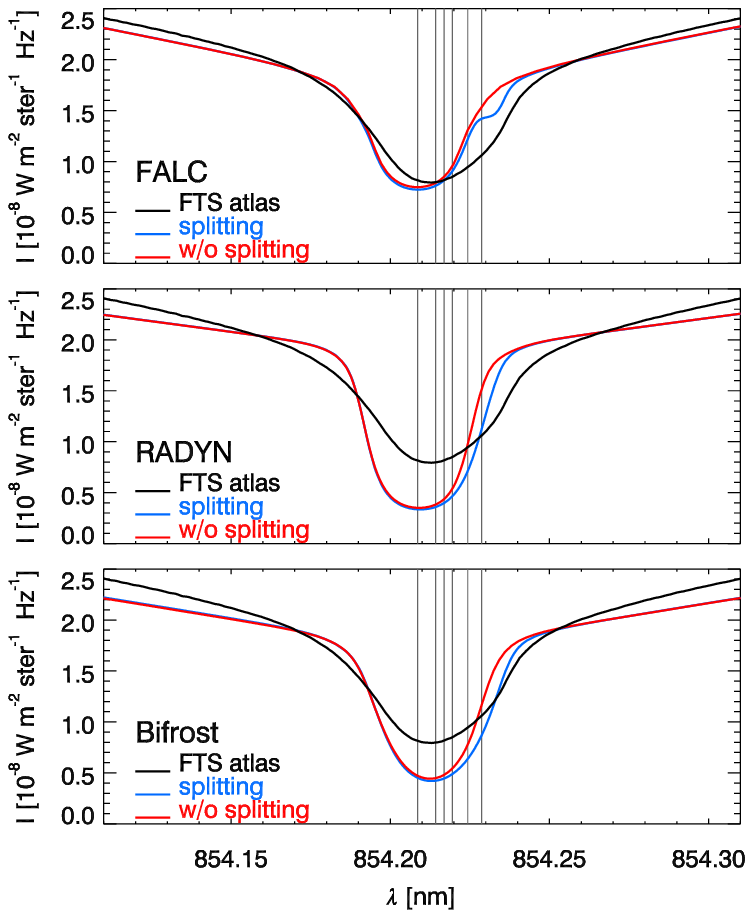}
   \includegraphics[width=\columnwidth]{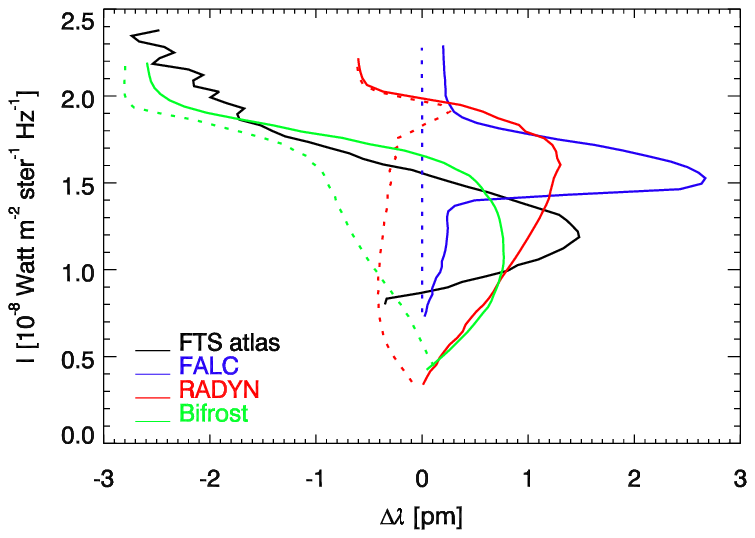}
  \caption{{\it Upper three panels:} comparison of the observed atlas profile (black) of the \Cair\ line with synthetic profiles computed from three different atmosphere models with isotopic splitting (blue) and without splitting (red). The vertical grey lines indicate the wavelength of the transition for each isotope (see Table~\ref{table:ca-table}). {\it Bottom panel:} bisectors of the \Cair\ line as determined from the atlas profile and the synthetic profiles from the atmosphere models. Solid curves are with splitting, dotted curves without splitting. The zero point of the $\Delta \lambda$-scale is the wavelength of the minimum intensity of each line profile.
  \label{fig:profiles}}
 \end{figure}

In Figure~\ref{fig:profiles} we show the vertically emergent profiles
of the \Cair\ line from our model atmospheres with and without
isotopic splitting and compare to the observed quiet-sun atlas profile
from
\citet{1984SoPh...90..205N}.
The profiles for the \radyn\ simulation are time averages. The Bifrost
profiles are spatial averages over the simulation box.

The synthetic profiles 
\edt{appear}
 all nearly symmetric when isotopic
splitting is ignored, and are narrower than the atlas. With the
inclusion of the isotopic splitting the profiles become wider and
asymmetric, also in the case of the static FALC atmosphere. The FALC
profile only show asymmetry in the upper part of the line core. This
lack of deep-core asymmetry is caused by the chromospheric temperature
rise that masks the presence of the heavier isotopes. The \radyn\ and
Bifrost profiles show a modest asymmetry that is present in the whole
line core. None of the models reproduces the observed core width or
intensity. \edt{The too small core width indicates insufficiently strong vertical motions
\citep[\cf][]{2009ApJ...694L.128L},
the too low core intensity is mainly caused by a too low temperature in the model chromospheres.}

Inspection of the 849.8 and 866.2~nm lines shows that those
lines behave the same: the line cores are \edt{nearly} symmetric without isotopic
splitting but show \edt{a strong} asymmetry when the splitting is included.

We computed the bisector of all line profiles. We define the bisector
$b(I)$ as function of the intensity $I$ as \be b(I) = \frac{1}{2}
\left( \lambda_\mathrm{red}(I) + \lambda_\mathrm{blue}(I) \right) -
\lambda_\mathrm{min}, \ee with $\lambda_\mathrm{red}(I)$ the
wavelength on the red side of the profile minimum where the intensity
is $I$, and similarly for $\lambda_\mathrm{blue}(I)$, the quantity
$\lambda_\mathrm{min}$ is the wavelength of the profile minimum. The
bisector is by construction only sensitive to the line asymmetry, and
not to the line width.  By defining the zero point to be the profile
minimum we allow direct comparison of the bisector amplitude and avoid
uncertainties in the absolute wavelength calibration of the atlas
profile.

We display the resulting bisectors in the bottom panel of
Figure~\ref{fig:profiles}. The atlas profile shows the observed
quiet-sun inverse-C shape. The models do not show this shape when
splitting is ignored. Note that we obtain the same bisector for the
\radyn\ model ignoring splitting as
\citet{2006ApJ...639..516U}. 

This inverse-C shape is qualitatively reproduced by the models when
isotopic splitting is included.  FALC has near-zero amplitude in the
deep core and a sudden increase of the amplitude in the upper core
beyond the observed amplitude. It does not display the turn towards
the blue at larger intensity as in the atlas profile. \radyn\ exhibits
the correct bisector amplitude, but not the turn towards the blue. The
Bifrost model does not reach the observed bisector amplitude, but
reproduces the blueward turn. \edt{As Bifrost is the only model with 
a convection zone, we speculate that this blueward turn is caused 
by overturning convection in the upper photosphere.}

\subsection{Observations}

To support our theoretical results, we explored a \Cair\ dataset
acquired at the SST with the CRISP instrument. The images were taken
at solar disk center in quiet-Sun on the 2013-07-14 at 10:38~UT. The
line was observed with a sampling of 5.6~pm in the line core up to
$\Delta\lambda=\pm 42$~pm. In the wings, the sampling was coarser.  At
854.2~nm, CRISP has a spectral resolution of $\delta \lambda=11.1$~pm,
thus our profiles are critically sampled close to line center.
The data have been processed as described in \citet{2013A&A...556A.115D}.

\citet{holemsc} analyzed a similar dataset, trying to associate the
inverse-C shaped bisector with specific solar features or
dynamics. The author detected the inverse-C shape in spatially
resolved profiles, \edt{and concluded the bisector shape is thus not an effect of averaging
 different profiles that individually do not show an inverse-C bisector.}

\begin{figure}
  \includegraphics[width=\columnwidth]{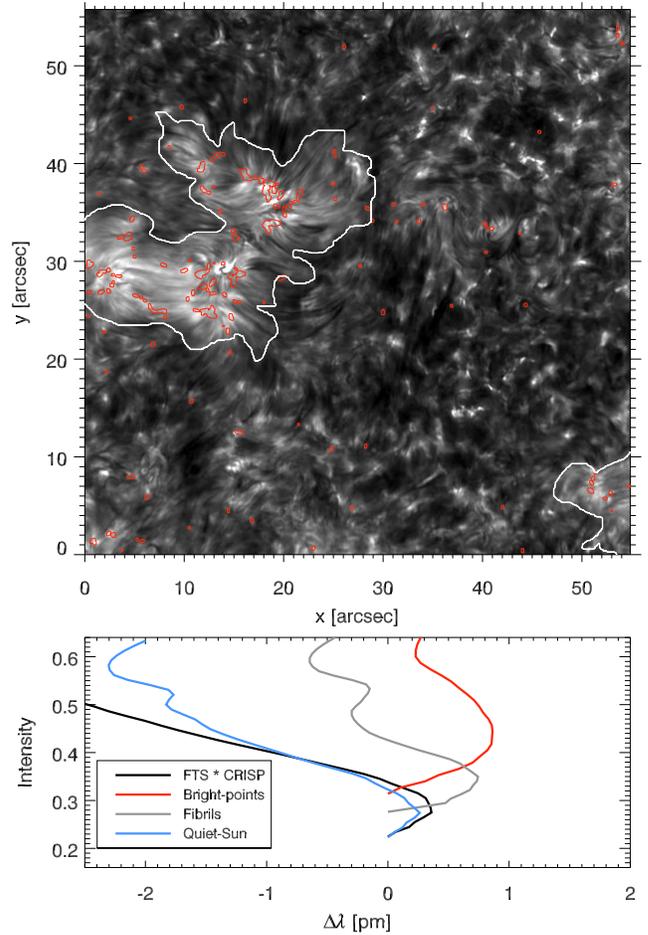}
  \caption{SST observations in the \Cair\ line. \emph{Top:} line core image. Masks enclosing bright-points (red) and fibrils (white) are indicated with contours. \emph{Bottom:} bisectors computed from spatially averaged line profiles, using the masks indicated in the top panel, for fibrils (gray), quiet-Sun (blue) and bright-points (red). The bisector from the FTS atlas convolved with the CRISP spectral profile is shown in black.
  \label{fig:sst}}
 \end{figure}

We divided the field-of-view in our data in three regions:
bright-points, chromospheric fibrils and the rest of the quiet
Sun. Then we computed spatially averaged line profiles within each
region, and calculated the bisectors as shown in
Figure~\ref{fig:sst}. We find that
inverse-C shaped bisectors are present in spectra from fibrils,
bright-points and quiet-Sun, \edt{but only the quiet-Sun bisector shows 
the blueward turn at higher intensities.} 
The bisector amplitudes in each region
are of the same order of magnitude to that from the FTS
atlas. Isotopic splitting naturally explains the ubiquitous presence
of the asymmetry, despite the variation in thermodynamical
properties. 

\edt{Note that the bisector amplitude is larger in fibrils and bright points than in quiet Sun, and the blueward turn is suppressed. This is in agreement with 
\citet{2013ApJ...764..153P}
who found the same behaviour in circumfacular regions compared to the quiet Sun.
}


\subsection{Influence on inversions}

Inversions of solar observations in the \Cair\ line have so far
ignored isotopic splitting
\citep[\eg][]{2001A&A...366..686T,2007ApJ...670..885P,2013A&A...556A.115D}. 
The additional line width and asymmetry caused by the minority
isotopes will then typically be fitted with a velocity gradient. We
investigated this effect by inverting the synthetic full Stokes vector
including isotopic splitting computed from a column from the Bifrost
simulation taken from a magnetic element. In order to mimic real
observing conditions we convolved the profile with the CRISP spectral
resolution and resampled it at the CRISP critical sampling of 5.6~pm.

The inversions were performed with the\nicole\ code in non-LTE 
\citep{2000ApJ...530..977S}, 
using an inversion strategy similar to %
\citet{2012A&A...543A..34D}.%
\nicole\ iteratively modifies the physical parameters in a 1D model
atmosphere to reproduce observed full-Stokes profiles. 

\begin{figure}
  \includegraphics[width=\columnwidth]{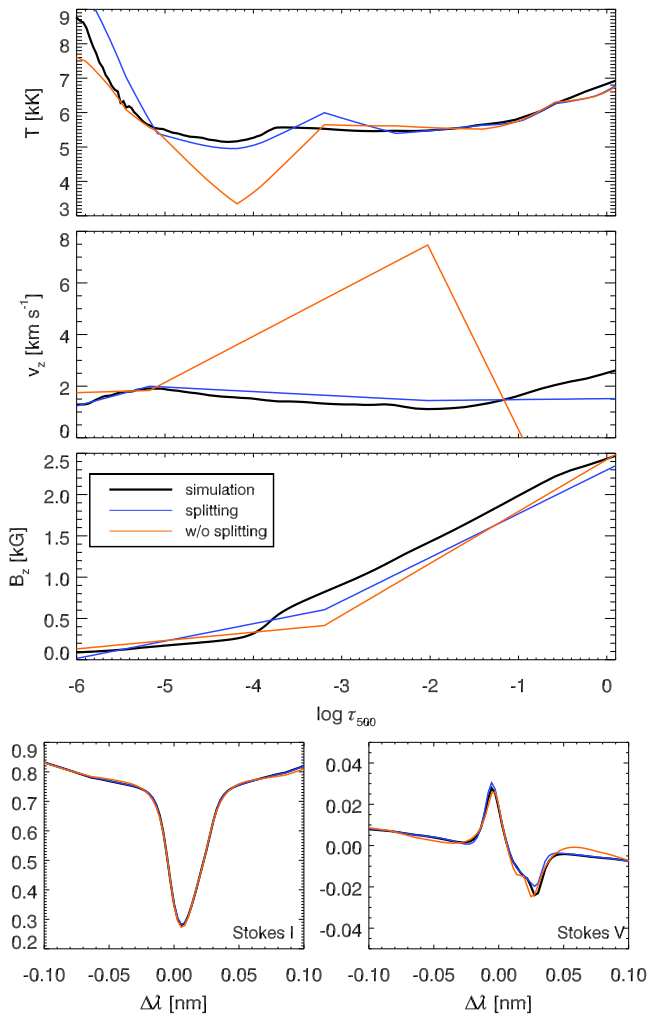}
  \caption{Comparison between the original Bifrost atmosphere (black) and the inverted quantities including isotopic splitting (blue) and without splitting (red). The three top rows show temperature, vertical velocity and the vertical magnetic field. The bottom row shows the fits for Stokes~$I$ and $V$ using the same color coding.
  \label{fig:inv}}
 \end{figure}

The inversion was performed twice. Once assuming all calcium is $^{40}$Ca, and once using an approximation to include overlapping lines by 
\citet{1986UppOR..33.....C}.
In this approximation the \Cair\ absorption coefficient is a weighted
sum of Voigt profiles, each centered at the rest-wavelength of the
line for each isotope. The weight is proportional to the relative
abundance of each isotope. This is an approximation because the ratio
of the level populations between different isotopes is not constant in
the atmosphere. This method is not as accurate as treating each
isotope separately, but it is much faster, and test calculations show
that the differences in the resulting line profiles are
insignificant. The inversion employed 11 temperature nodes and 4 nodes
for the vertical velocity and 3 nodes for the vertical magnetic
field. The results are shown in Figure~\ref{fig:inv}.

The inversion including isotopic splitting recovers Stokes~$I$ and the
atmospheric parameters well, but it does not quite fit the amplitude
of the Stokes~$V$ peak on the red side of the line core.

If isotopic splitting is ignored, both Stokes $I$ and $V$ are fitted
slightly worse. In order to fit the profile asymmetry, the inversion
without splitting introduces a velocity field with up to 6 km~s$^{-1}$
difference to the true value and adjusts the temperature up to
2~kK. The resulting inferred atmosphere does not resemble the original
atmosphere.

The Zeeman response in the \Cair\ line is usually in the weak-field regime
 \citep[$B < 2500$~G,  see][]{2013A&A...556A.115D}. 
Therefore, if the inversion is able to fit Stokes~$I$, then the magnetic field is retrieved accurately irrespective of whether isotopic splitting is included in the inversion. This is the case in our computation. But if the inversion cannot reproduce Stokes~$I$, errors can also appear in the inferred magnetic field.

\begin{figure}
  \includegraphics[width=\columnwidth]{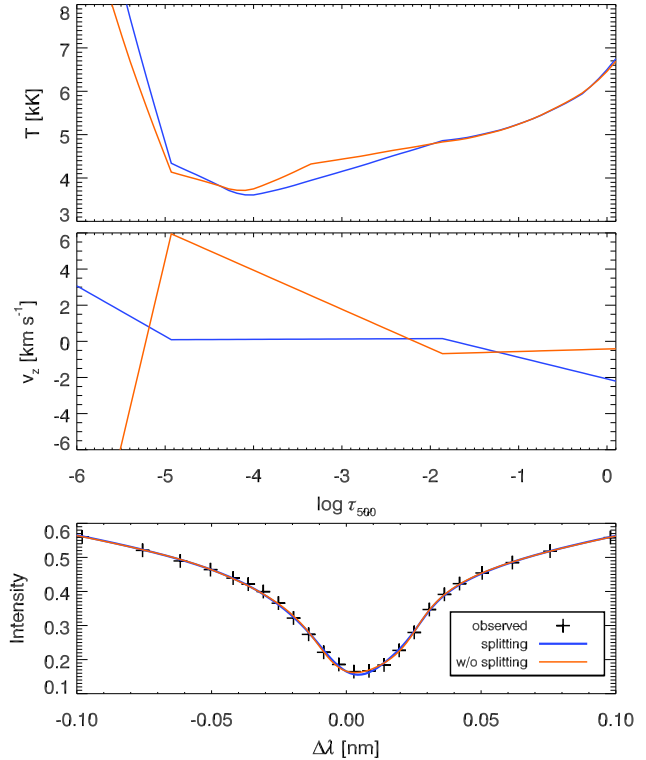}
  \caption{Inversion of an observed quiet-Sun profile from our
    observations, including isotopic splitting (blue) and without
    isotopic splitting (red). From top to bottom, the panels show
    temperature, line-of-sight velocity and the line profile. The
    observed profile is indicated with black crosses in the bottom
    panel.
  \label{fig:sst_inv}}
 \end{figure}

We also inverted the intensity profile from a row of pixels from the
observations (512 in total) using 7 nodes in temperature and 4 nodes
in vertical velocity. These profiles were also inverted twice: with
and without isotopic splitting. Figure~\ref{fig:sst_inv} shows the
results for a pixel with a clearly asymmetric line profile. The line
profile is fitted equally well with both methods. However, the
inversion without splitting introduces a velocity field in order to
fit the line shape, similar to the inversion of the synthetic
profile. Both inversions provide very similar estimates for the
temperature as a function of height, in contrast to the synthetic
data. The difference in inferred velocity structure is common: 62\% of
the pixels we inverted with and without splitting show a difference
larger than 2~km~s$^{-1}$.

Finally, we note that the inversion without splitting needs four
velocity nodes in order to fit the profile. The inversion including
splitting reaches a good fit using only two nodes.

\section{Discussion and conclusions} \label{sec:conclusions}

The various isotopes of \Ca\ have slightly different energies for the
3d and 4p levels of \CaII. This results in isotopic splitting of the
infrared triplet lines. This is of importance because the \Cair\ line
is an often-used diagnostic of the solar chromosphere.

We investigated this splitting in a static 1D and dynamic 1D and 3D
models of the solar atmosphere. All employed models produce an
asymmetric line core with an inverse-C shaped bisector. These effects
are not present when ignoring the minority isotopes. No model
reproduces the observed quiet sun spectrum quantitatively, but all
models agree qualitatively. We therefore conclude that the \Cair\ line
asymmetry and bisector shape is mainly caused by isotopic splitting,
and not by effects caused by velocity fields as has been suggested
earlier
\citep{2006ApJ...639..516U}.

The exact bisector shape and amplitude \edt{in our models} depend on the atmospheric
structure. \edt{In our observations we find a markedly different bisector shape between 
quiet Sun and regions with stronger magnetic field, although both show the inverse-C shape. 
This agrees with the findings of
\citet{2013ApJ...764..153P}
} 
and the variation of the bisector shape with the solar cycle
\citep{2011ApJ...736..114P}.

In addition we investigated the effect of isotopic splitting on the
inversion of line profiles. We found that ignoring isotopic splitting
in a test inversion of a synthetic line profile leads to an inferred
atmospheric structure that contains spurious velocity gradients and an
incorrect temperature variation in order to fit the line-core
asymmetry. Inversions of observed line profile\edt{s} indicate similar
differences in inferred velocities with and without the inclusion of
isotopic splitting.  Modelling splitting by adding the line profiles
of the different isotopes together largely eliminates these biases,
without additional computational cost.

We recommend that all future inversions of the \CaII\ infrared lines
include the effects of isotopic splitting.

\begin{acknowledgements}
   This research was supported by the Research Council of Norway
   through the grant ``Solar Atmospheric Modeling'', from the European
   Research Council under the European Union's Seventh Framework
   Programme \mbox{(FP7/2007-2013)} / ERC Grant agreement no. 291058,
   and through grants of computing time from the Programme for
   Supercomputing of the Research Council of Norway. The Swedish 1-m Solar Telescope is operated on the island of La Palma by the Institute for Solar Physics of Stockholm University in the Spanish Observatorio del Roque de los Muchachos of the Instituto de Astrof'sica de Canarias. We thank Luc Rouppe van der Voort for illuminating discussions, and Andrew McWilliam for pointing out the hyperfine splitting of $^{43}$Ca.
\end{acknowledgements}


\end{document}